\documentclass[twocolumn]{aastex61}

\newcommand{\HI}{\ion{H}{1}}
\newcommand{\kms}{\ensuremath{\rm{km\,s^{-1}}}}
\newcommand{\mlim}{\ensuremath{M_{HI}^{lim}}}
\newcommand{\msun}{\ensuremath{M_{\odot}}}
\newcommand{\ml}{\ensuremath{M_{HI}^{lim}/L_V}}
\newcommand{\mlsun}{\ensuremath{M_\odot/L_\odot}}
\usepackage[]{threeparttable}
\usepackage{float}
\usepackage[caption = false]{subfig}
\usepackage{natbib}

\shorttitle{Sgr~II, Ret~II, Phe~II and Tuc~III}
\shortauthors{MUTLU-PAKD\.{I}L et al.}

\begin{document}

\title{A Deeper Look at the New Milky Way Satellites: Sagittarius~II, Reticulum~II, Phoenix~II, and Tucana~III\footnote{This paper includes data gathered with the 6.5~m Magellan Telescopes at Las Campanas Observatory, Chile.}}
\correspondingauthor{Bur\c{c}in Mutlu-Pakdil}
\email{bmutlupakdil@as.arizona.edu}

\author{BUR\c{C}\.{I}N MUTLU-PAKD\.{I}L}
\affil{Department of Astronomy/Steward Observatory, 933 North Cherry Avenue, Rm. N204, Tucson, AZ 85721-0065, USA}

\author{DAVID J. SAND}
\affil{Department of Astronomy/Steward Observatory, 933 North Cherry Avenue, Rm. N204, Tucson, AZ 85721-0065, USA}

\author{JEFFREY L. CARLIN}
\affil{LSST, 950 North Cherry Avenue, Tucson, AZ 85719, USA}

\author{KRISTINE SPEKKENS}
\affil{Department of Physics, Engineering Physics and Astronomy, Queen’s University, Kingston, Ontario, Canada}
\affil{Department of Physics, Royal Military College of Canada, P. O. Box 17000, Station Forces, Kingston, ON K7L 7B4, Canada}

\author{NELSON CALDWELL}
\affil{Harvard-Smithsonian Center for Astrophysics, 60 Garden Street, Cambridge, MA 02138}

\author{DENIJA CRNOJEVI\'{C}}
\affil{Texas Tech University, Physics \& Astronomy Department, Box 41051, Lubbock, TX 79409-1051, USA}

\author{ALLISON K. HUGHES}
\affil{University of Arizona, Department of Physics, 1118 E. Fourth Street, Tucson, AZ 85721}

\author{BETH WILLMAN}
\affil{LSST and Steward Observatory, 933 North Cherry Avenue, Tucson, AZ 85721, USA}

\author{DENNIS ZARITSKY}
\affil{Department of Astronomy/Steward Observatory, 933 North Cherry Avenue, Rm. N204, Tucson, AZ 85721-0065, USA}
 
\begin{abstract}
We present deep Magellan/Megacam stellar photometry of four recently discovered faint Milky Way satellites:
Sagittarius~II (Sgr~II), Reticulum~II (Ret~II), Phoenix~II (Phe~II), and Tucana~III (Tuc~III). 
Our photometry reaches $\sim$2-3 magnitudes deeper than the discovery data, allowing us to revisit the properties
of these new objects (e.g., distance, structural properties, luminosity measurements, and signs
of tidal disturbance). The satellite color-magnitude diagrams show that they are all old ($\sim$13.5 Gyr) 
and metal-poor ([Fe/H]$\lesssim-2.2$). Sgr~II is particularly interesting as it sits in an intermediate 
position between the loci of dwarf galaxies and globular clusters in the size-luminosity plane. The ensemble of its structural 
parameters is more consistent with a globular cluster classification, indicating that Sgr~II is the most extended globular cluster 
in its luminosity range. The other three satellites land directly on the locus defined by Milky Way ultra-faint dwarf galaxies 
of similar luminosity. Ret~II is the most elongated nearby dwarf galaxy currently known for its luminosity range.  
Our structural parameters for Phe~II and Tuc~III suggest that they are both dwarf galaxies. 
Tuc~III is known to be associated with a stellar stream, which is clearly visible in our matched-filter stellar density map. 
The other satellites do not show any clear evidence of tidal stripping in the form of extensions or 
distortions. Finally, we also use archival HI data to place limits on the gas content of each object.
\end{abstract}


\section{INTRODUCTION} \label{sec:intro}

Ultra-faint dwarf galaxies are the smallest, the most dark-matter dominated, and the least chemically 
enriched stellar systems in the Universe. They are important targets for understanding the physics of 
dark matter and galaxy formation on the smallest scales. The details of their nature provide crucial 
empirical input for verifying formation scenarios of the Milky Way (MW).

Many new MW satellites have been discovered in the last few years \citep[e.g.,][]{Balbinot2013,Belokurov2014,Laevens2014,Bechtol2015,Drlica2015,Kim2015,Koposov2015a,Laevens2015,Martin2015,Drlica2016,Torrealba2016,Koposov2017,Torrealba18,Koposov18}, but some of these new discoveries have been called into question by recent deep imaging campaigns \citep{Conn2018a,Conn2018b} and most of these new objects are poorly constrained in terms of their stellar population, structural parameters, 
distance and luminosity. This paper aims to derive more accurate constraints on four ultra-faint
satellites $-$ Sagittarius~II (Sgr~II), Reticulum~II (Ret~II), Phoenix~II (Phe~II), and Tucana~III (Tuc~III)
$-$ by analyzing deep photometric observations.

Sgr~II was discovered in the PanSTARRS (PS1) survey by \citet{Laevens2015}. It is especially interesting
as it is either the most compact dwarf galaxy or the most extended globular cluster in its luminosity-size
range. Ret~II and Phe~II were discovered in the first-year Dark Energy Survey (DES) independently 
by \citet{Bechtol2015} and \citet{Koposov2015a}. As with several recently-discovered
satellites in DES, the derived structural parameters differ between these studies. Deeper photometry 
is necessary to resolve this discrepancy. Tuc~III was discovered in the second-year DES data by 
\citet{Drlica2015}. Spectroscopic observations have been unable to conclusively 
determine its dynamical status and dark matter content \citep{Simon2017}, and it has a stellar stream 
extending at least $\pm2^{\circ}$ from its core \citep{Drlica2015,Shipp2018},
which strongly influences the measurements of its size and shape from the discovery DECam data. 

Dark matter may annihilate to produce gamma rays \citep[e.g.,][]{Gunn1978,Bergstrom1988,Baltz2008}. 
The J-factor is a measure of the strength of this predicted signal and can be
estimated using stellar kinematics. Due to their large dark matter content, relative proximity, and 
low astrophysical foregrounds, the MW dwarf galaxies are promising targets for indirect dark matter searches. 
However, it can be hard to obtain reliable J-factor estimates due to the difficulty of obtaining spectroscopy 
on these faint objects. Recently, \citet{Pace2018} derived a scaling relation to estimate the J-factor
based on the observable properties, such as half-light radius, without the full dynamical analysis. 
This makes deeper photometric data particularly important to precisely measure the structural parameters 
for these galaxies. We note that spectroscopic observations have only been published for Ret~II and Tuc~III, 
not Sgr~II and Phe~II.
Spectroscopic data of Tuc~III only provides an upper limit on its J-factor \citep{Simon2017}. 
A tentative and controversial gamma-ray detection from Ret~II \citep{Geringer2015,Hooper2015} 
demands detailed measurement of all of its physical parameters, and an evaluation of its dynamical status.

It is also possible to search for signs of tidal disruption via deep wide-field imaging
\citep[e.g.,][]{Coleman07,Sand2009,Munoz2010,Sand2012,Roderick15,Carlin2017}, and this  can be an important 
probe of the dynamical status of a satellite.  Follow-up observations of these signs of structural disturbance 
have led to additional evidence of disruption in the form of extended distributions of RR Lyrae stars 
\citep[e.g.][]{Garling18} and velocity gradients \citep{Aden09,Collins17} in some new MW satellites.

An outline of the paper follows. In Section~\ref{sec:obs}, we describe the observations, photometry, and
calibration of our data. Section~\ref{sec:analysis} details our photometric analysis, including new distance 
measurements, structural parameter measurements, and a search for signs of extended/disturbed structure. We also 
use archival HI data to place upper limits on the gas content of each object. In Section~\ref{sec:disc} we discuss 
the individual objects in turn. Finally, we summarize and conclude in Section~\ref{sec:conc}.

\section{Observations and Data Reduction} \label{sec:obs}

\setcounter{table}{0}
\begin{table*}[ht!]
\renewcommand{\thetable}{\arabic{table}}
\centering
\caption{Summary of Observations and Field Completeness.} \label{tab:obs}
\begin{tabular}{ccccccc}
\tablewidth{0pt}
\hline
\hline
Dwarf Name & UT Date &  Filter & Exp & PSF FWHM & 50\% & 90\% \\
{}         & {} &  {} & (s) & (arcsec) & (mag) & (mag) \\
\hline
Sgr~II &  2015 Oct 12 &  $g$ & 6x300 & 0.8 & 26.23 & 24.93    \\
{}     &  2015 Oct 12 &  $r$ & 6x300 & 0.8 & 25.81 & 24.13    \\
Ret~II &  2017 Oct 17 &  $g$ & 8x300 & 1.1 & 25.97 & 24.84    \\
{}     &  2015 Oct 13 &  $r$ &14x300 & 0.8 & 25.55 & 24.02    \\
Phe~II &  2015 Oct 12 &  $g$ & 5x300 & 0.6 & 26.52 & 25.53    \\
{}     &  2015 Oct 12 &  $r$ & 6x300 & 0.6 & 26.05 & 24.79    \\
Tuc~III & 2015 Oct 12 &  $g$ & 9x300 & 0.7 & 26.90 & 25.33    \\
{}      & 2015 Oct 12 &  $r$ & 7x300 & 0.6 & 26.37 & 24.68    \\
\hline
\end{tabular}
\end{table*}

We observed our targets with Megacam \citep{McLeod2000} at the $f$/5 focus 
at the Magellan Clay telescope in the $g$ and $r$ bands. The data were mostly 
taken during a single run on October 12-13, 2015, and was supplemented by some 
observations taken on October 17, 2017. Data from this program on Eridanus~II 
was previously presented in \citet{Crnojevic2016}, although we will not discuss 
those observations further here. Magellan/Megacam has 36 CCDs, 
each with 2048$\times$2048 pixels at 0.08\arcsec/pixel (which were binned 2$\times$2),
for a total field of view (FoV) of $24\arcmin \times 24\arcmin$. We obtained 300 s individual exposures in $g$
and $r$ bands, and small dithers were used to cover the chip gaps in the final stacks.
We reduced the data using the Megacam pipeline developed at the Harvard-Smithsonian Center for Astrophysics 
by M. Conroy, J. Roll, and B. McLeod, including detrending the data, performing astrometry, and 
stacking the individual dithered frames. Astrometric solutions for each science exposure
were derived using Two Micron All Sky Survey catalog \citep[2MASS,][]{Skrutskie2006}. Typical 
residuals for the matches to the 2MASS catalog are approximately 120 mas \citep{McLeod2015}.

We perform point-spread function fitting photometry on the final image stacks, using the 
DAOPHOTII/ ALLSTAR package \citep{Stetson1994} and following the same methodology described in
\citet{Sand2009}. In short, we use a quadratically varying point-spread function (PSF) 
across the field to determine our PSF model. We run {\sc Allstar} twice: first on the final stacked
image, and then on the first round's star-subtracted image, in order to recover fainter sources. 
We remove objects that are not point sources by culling our {\sc Allstar} catalogs of outliers 
in $\chi$ versus magnitude, magnitude error versus magnitude, and sharpness versus 
magnitude space. We positionally match our source catalogs derived from $g$ and $r$ filters 
with a maximum match radius of 0.5\arcsec, and create our final catalogs by only keeping those 
point sources detected in both bands.

We calibrate the photometry for Sgr~II by matching to the PS1 survey \citep{Chambers2016,Flewelling2016,Magnier2016}.
The Ret~II, Phe~II and Tuc~III fields were calibrated by matching to the DES DR1 catalog \citep{DEScollaboration18}. 
We note that the DES did not cover the Sgr~II field, and the PS1 did not cover the others.
A zeropoint and color term were fit for all fields and filters. We use all stars within the FoV with 
$17.5 < g < 20.5$ and $17.5 < r < 20.5$. A final overall scatter about the best-fit zero point is $\lesssim0.01$ mag 
in all of our photometric bands. We correct for MW extinction on a star by star 
basis using the \citet{Schlegel1998} reddening maps with the coefficients from \citet{Schlafly2011}, which were evaluated according to an \citet{Fitzpatrick1999} reddening law with normalization $N=0.78$. Specifically,  
we adopted the coefficients of 3.172 and 2.271 for PS1 $g$ and $r$, 3.237 and 2.176 for DES $g$ and $r$ to use with 
$E(B-V)$. The extinction-corrected photometry is used throughout this work.

To determine our photometric errors and completeness as a function of magnitude and color, 
we perform a series of artificial star tests with the DAOPHOT routine ADDSTAR. Similar to 
\citet{Sand2012}, we place artificial stars into our images on a regular grid ($10\mbox{--}20$ 
times the image FWHM). We assign the $r$ magnitude of the artificial stars randomly from 18 to 
29 mag with an exponentially increasing probability toward fainter magnitudes, and the $g$ magnitude 
is then randomly selected based on the $g-r$ color over the range $-0.5 \mbox{--} 1.5$ mag, with  
uniform probability. Ten iterations are performed on each field for a total of $\sim$100,000
artificial stars each. These images are then photometered in the same way as the unaltered 
image stacks and the same stellar selection criteria on $\chi$, magnitude, magnitude error, and sharpness were 
all applied to the artificial star catalogs in order to determine our completeness and average magnitude uncertainties.

We present an observation log and our completeness limits in Table~\ref{tab:obs}.  
Our photometry reaches $\sim$2-3 mag deeper than the original discovery data for each object. 
Tables~\ref{tab:sag2}-\ref{tab:tuc3} present our full 
Sgr~II, Ret~II, Phe~II, and Tuc~III catalogs, which
include the calibrated magnitudes (uncorrected for extinction) along with their DAOPHOT uncertainty,
as well as the Galactic extinction values derived for each star \citep{Schlafly2011}. 

\section{ANALYSIS}\label{sec:analysis}

\subsection{Color-Magnitude Diagrams} \label{subsec:cmd}

\setcounter{table}{1}
\begin{table*}[tb!]
\renewcommand{\thetable}{\arabic{table}}
\caption{Sgr~II Photometry in the PS1 photometric system.} \label{tab:sag2}
\begin{minipage}[b]{\linewidth}\centering
\begin{tabular}{ccccccccc}
\tablewidth{0pt}
\hline
\hline
Star No. & $\alpha$      &  $\delta$      & $g$     & $\delta g$ & $A_{g}$ & $r$     & $\delta r$ & $A_{r}$ \\
{}       & (deg J2000.0) &  (deg J2000.0) & (mag) & (mag)     & (mag)   & (mag) & (mag)     & (mag)   \\
\hline
0 &    297.93544 &     -22.131170   &   18.87  &  0.01  &  0.35  &   17.88  &  0.01  &  0.25\\
1 &    297.93544 &     -22.219629   &   18.31  &  0.01  &  0.35  &   17.80  &  0.01  &  0.25\\
2 &    297.93550 &     -22.031850   &   19.81  &  0.01  &  0.36  &   19.13  &  0.02  &  0.26\\
3 &    297.93562 &     -22.140471   &   19.61  &  0.01  &  0.35  &   19.12  &  0.01  &  0.25\\
4 &    297.93651 &     -22.144029   &   19.69  &  0.01  &  0.35  &   19.15  &  0.01  &  0.25\\
\hline
\end{tabular}
   \begin{tablenotes}
      \small
      \item (This table is available in its entirety in a machine-readable form in the online journal. A portion
      is shown here for guidance regarding its form and content.)
    \end{tablenotes}
\caption{Ret~II Photometry in the DES photometric system.} \label{tab:ret2}
\begin{tabular}{ccccccccc}
\tablewidth{0pt}
\hline
\hline
Star No. & $\alpha$      &  $\delta$      & $g$     & $\delta g$ & $A_{g}$ & $r$     & $\delta r$ & $A_{r}$ \\
{}       & (deg J2000.0) &  (deg J2000.0) & (mag) & (mag)     & (mag)   & (mag) & (mag)     & (mag)   \\
\hline
0 &    53.565994  &  -54.247698  &   18.04  &  0.01  &  0.06 &   17.03 &   0.01 &    0.04\\
1 &    53.566061  &  -54.247761  &   18.26  &  0.02  &  0.06 &   17.01 &   0.01 &    0.04\\
2 &    53.576225  &  -54.168327  &   19.47  &  0.02  &  0.05 &   17.38 &   0.01 &    0.03\\
3 &    53.576270  &  -54.168395  &   18.32  &  0.01  &  0.05 &   17.45 &   0.01 &    0.03\\
4 &    53.578226  &  -53.952364  &   18.11  &  0.01  &  0.05 &   15.54 &   0.01 &    0.03\\
\hline
\end{tabular}
   \begin{tablenotes}
      \small
      \item (This table is available in its entirety in a machine-readable form in the online journal. A portion
      is shown here for guidance regarding its form and content.)
    \end{tablenotes}
\caption{Phe~II Photometry in the DES photometric system.} \label{tab:phe2}
\begin{tabular}{ccccccccc}
\tablewidth{0pt}
\hline
\hline
Star No. & $\alpha$      &  $\delta$      & $g$     & $\delta g$ & $A_{g}$ & $r$  & $\delta r$ & $A_{r}$ \\
{}       & (deg J2000.0) &  (deg J2000.0) & (mag) & (mag)     & (mag)   & (mag) & (mag)     & (mag)   \\
\hline
0 &   354.65054  &  -54.405307  &  19.50  &  0.01  &   0.04  &  18.05 &   0.01  &   0.03\\
1 &   354.65970  &  -54.537558  &  19.93  &  0.01  &   0.04  &  19.00 &   0.01  &   0.03\\
2 &   354.66431  &  -54.542757  &  18.77  &  0.01  &   0.04  &  17.74 &   0.01  &   0.03\\
3 &   354.67117  &  -54.417531  &  18.26  &  0.01  &   0.04  &  17.72 &   0.01  &   0.02\\
4 &   354.68689  &  -54.386980  &  19.83  &  0.02  &   0.04  &  19.58 &   0.01  &   0.02\\
\hline
\end{tabular}
   \begin{tablenotes}
      \small
      \item (This table is available in its entirety in a machine-readable form in the online journal. A portion
      is shown here for guidance regarding its form and content.)
    \end{tablenotes}
\caption{Tuc~III Photometry in the DES photometric system.} \label{tab:tuc3}
\begin{tabular}{ccccccccc}
\tablewidth{0pt}
\hline
\hline
Star No. & $\alpha$      &  $\delta$      & $g$     & $\delta g$ & $A_{g}$ & $r$     & $\delta r$ & $A_{r}$ \\
{}       & (deg J2000.0) &  (deg J2000.0) & (mag) & (mag)     & (mag)   & (mag) & (mag)     & (mag)   \\
\hline
0 &    358.88627  & -59.427008  & 18.20 &   0.01 &  0.040   &  17.43  &  0.01  &   0.03\\
1 &    358.89465  & -59.422329  & 19.64 &   0.01 &  0.040   &  19.07  &  0.01  &   0.03\\
2 &    358.89886  & -59.650621  & 19.68 &   0.01 &  0.038   &  18.25  &  0.01  &   0.02\\
3 &    358.90786  & -59.681289  & 18.28 &   0.01 &  0.038   &  16.78  &  0.01  &   0.03\\
4 &    358.91664  & -59.438449  & 18.69 &   0.01 &  0.040   &  17.29  &  0.01  &   0.03\\
\hline
\end{tabular}
    \begin{tablenotes}
      \small
      \item (This table is available in its entirety in a machine-readable form in the online journal. A portion
      is shown here for guidance regarding its form and content.)
    \end{tablenotes}
\end{minipage}    
\end{table*}

\begin{figure}[ht!]
\centering
\includegraphics[width = 3.6in]{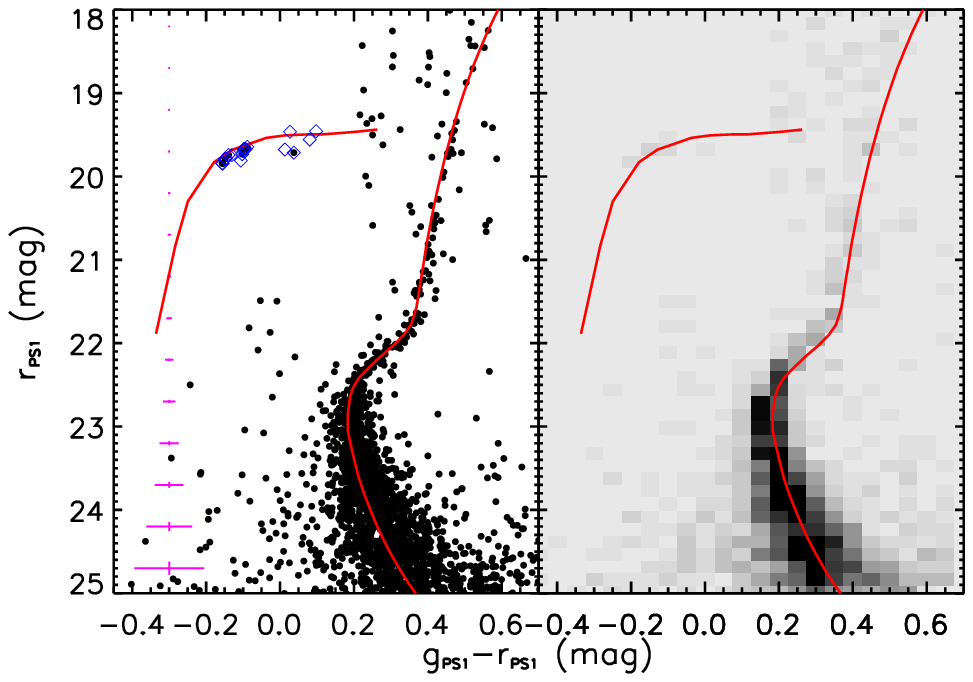} 
\caption{Left: CMD of Sgr~II, including stars within one half-light radius of its center. Magenta error 
bars show the color and magnitude uncertainties as a function of $r$ magnitude. Blue open diamonds are blue 
HB candidates within our FoV. There are a total of 19 candidates, 10 of which are within its half-light radius.  
Right: Background-subtracted binned Hess diagram of Sgr II for the same selected 
region shown in the left panel. Overplotted as a red line is the PS1 fiducial for M15 
from \citet{Bernard2014}.
}\label{fig:cmdsag2}
\end{figure}

\begin{figure*}[ht!]
\centering
\subfloat[Ret~II]{\includegraphics[width = 2.4in]{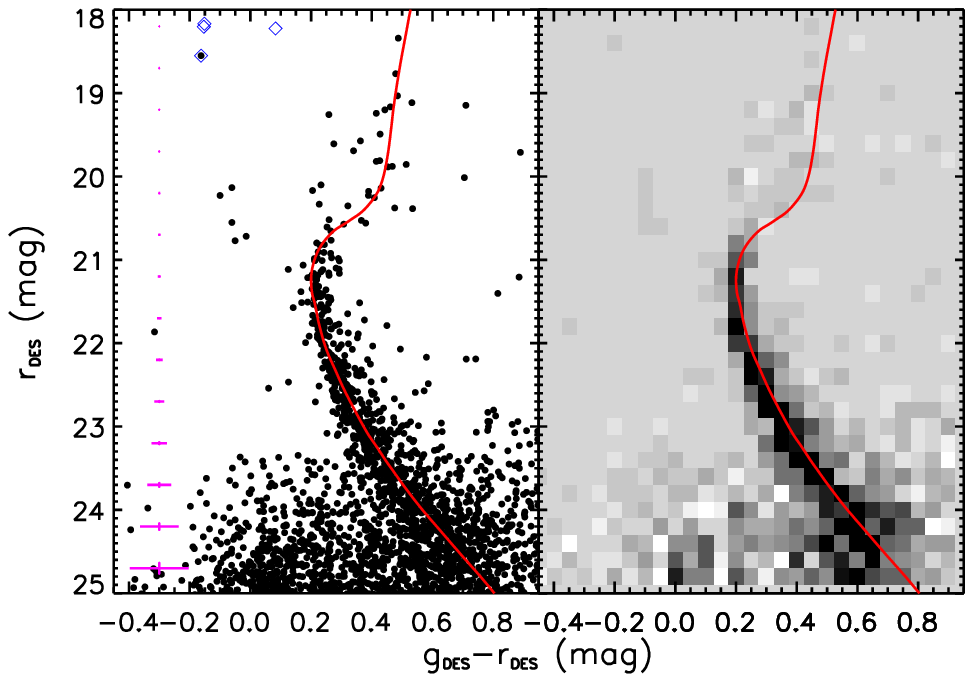}}
\subfloat[Phe~II]{\includegraphics[width = 2.4in]{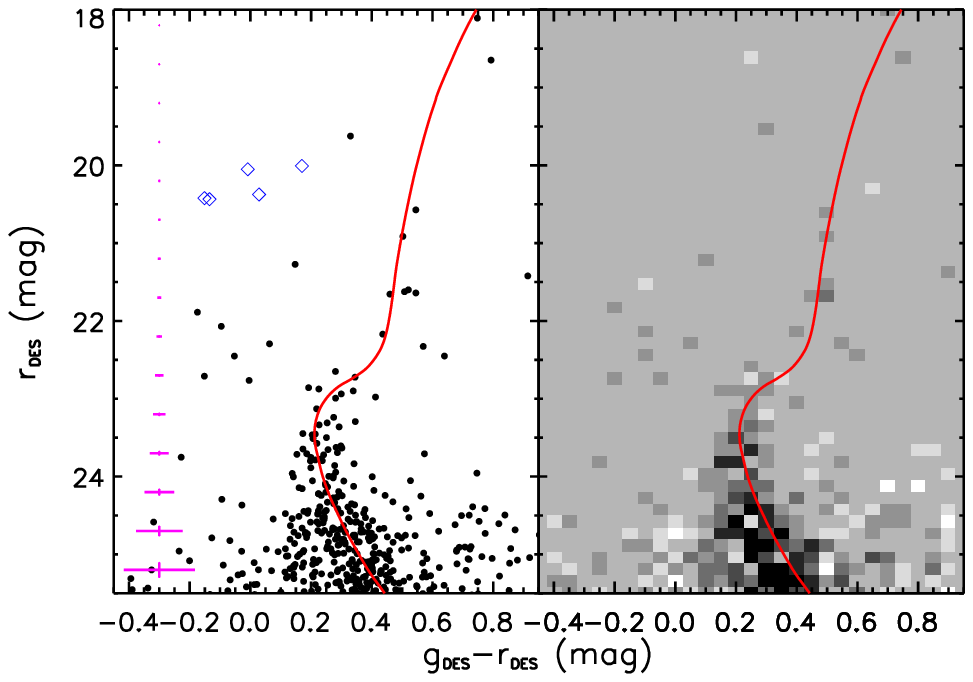}}
\subfloat[Tuc~III]{\includegraphics[width = 2.4in]{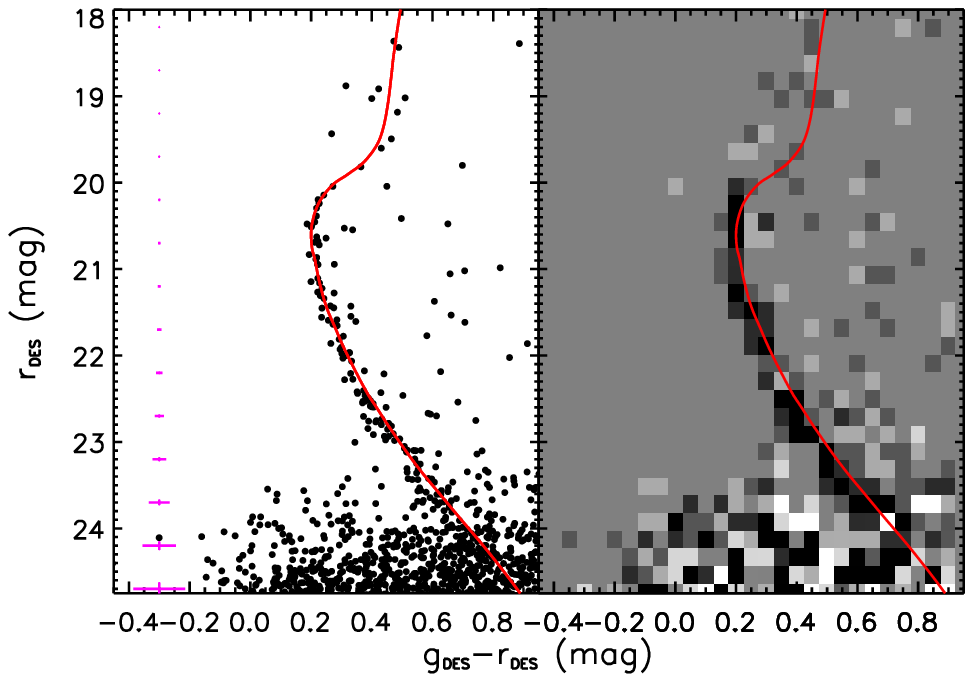}} 
\caption{CMDs and Hess diagrams of Ret~II (a), Phe~II (b) and Tuc~III (c), including stars within 
one half-light radius of their center. Magenta error bars show the color and magnitude uncertainties
as a function of $r$ magnitude. Blue open diamonds are blue HB candidates within our FoV.
Overplotted as a red line is a metal-poor Dartmouth isochrone with age 13.5 Gyr: [Fe/H]$=-2.4$ for Ret~II and Tuc~III, 
[Fe/H]$=-2.2$ for Phe~II.
}\label{fig:cmdhess}
\end{figure*}

Figures~\ref{fig:cmdsag2} (Sgr~II) and~\ref{fig:cmdhess} (Ret~II, Phe~II, Tuc~III) show the final color magnitude diagrams (CMDs), which include stars within one half-light radius of the center (see Section~\ref{subsec:Str}). 
Magenta error bars show the mean photometric errors determined from artificial stars. The error bars
are plotted at an arbitrary color for convenience. Blue open diamonds are the blue horizontal-branch star (HB) 
candidates in our FoV, which are selected within a filter with a color of $<0.2$ and a span of $\pm0.5$ mag, centered 
on the HB sequence of a metal-poor PARSEC isochrone \citep{Bressan2012}. 
Right panels show background-subtracted binned Hess diagrams, which encode the number density of stars in the 
selected region. The background is estimated from a field located outside a radius of 12\arcmin, 
well outside the body of each satellite. For Ret~II and Tuc~III, we use background fields that are northwest 
and southeast of the centroids because there should be little satellite contamination at these positions, given the position angle and high-ellipticity of Ret~II (see Table~\ref{tab:str}) and the orientation of Tuc~III stream (see Section~\ref{sec:tuciii}).
In Figure~\ref{fig:cmdhess}, overplotted as a red line is a metal-poor Dartmouth isochrone \citep{Dotter2008},
which corresponds to the best-fit Dartmouth isochrone in Section~\ref{subsec:dist}.
In Figure~\ref{fig:cmdsag2}, we use the PS1 fiducial for globular cluster NGC~7078 (M15) from \citet{Bernard2014}, 
which is a better fit for Sgr~II than any theoretical isochrone. Note that there is not any established DES fiducial
for globular clusters, and that is why we display theoretical isochrones in Figure~\ref{fig:cmdhess}.
We shift each isochrone and fiducial by the best-fit distance modulus that we derive in Section~\ref{subsec:dist}. 

\subsection{Distance}

We derive the distance to Sgr~II by comparing its CMD with empirical 
globular cluster fiducials and theoretical isochrones. We use four empirical fiducials determined
by \citet{Bernard2014} using the PS1 photometry: NGC~7078 (M15), NGC~6341 (M92), NGC~6205 (M13) 
and NGC~5272 (M3) with [Fe/H]$=$-2.42, -2.38, -1.60, and -1.50 \citep{Kraft2003}. We assume the 
distance modulus values of $m-M$=15.25 for M15, 14.75 for M92, 14.42 for M13, and 15.02 for M3 
\citep{Kraft2003}, with an uncertainty of 0.1 mag. Besides these four empirical fiducials, we also use the Dartmouth isochrones 
with [Fe/H]$=-2.20$, [$\alpha$/Fe]$=0.40$, and a 13.5 Gyr stellar population. 

\label{subsec:dist}
To determine the distance modulus of Sgr~II, we follow a very similar methodology as that described in 
\citet{Sand2009} (see also \citealt{Walsh2008}). We include all stars with $r <24$ mag within $r_{h}=1.58\arcmin$ of its 
center. Each fiducial is shifted through 0.025 mag intervals in $(m-M)$ from 18.0 to 20.0 mag. 
In each step, we count the number of stars consistent with the fiducial, taking into account 
photometric uncertainties. The selection region is defined by two selection boundaries on either side along 
the $g-r$ color axis at the typical color uncertainty at a given $r$ magnitude, as determined via our artificial star tests.
We also account for background stars by running the identical procedure 
in parallel over an appropriately scaled background region offset from Sgr~II, counting the number 
of stars consistent with the fiducial and then subtracting this number from that derived at the position of the dwarf.
We derive the best-fit distance modulus when the fiducial gives the maximum number of dwarf stars: 
19.250 for M15 (150 stars), 19.125 for M92 (170 stars), 18.675 for M13 (93 stars), 
18.700 for M3 (68 stars). Sgr~II's CMD is clearly more consistent with the metal-poor fiducials, 
i.e., M15 and M92. The Dartmouth isochrone gives $(m-M)$=19.225 mag with 112 stars. We use a 100 iteration
bootstrap analysis to determine the uncertainties on each fit. 

We also derive a distance modulus using the possible blue HBs of Sgr~II within our FoV (19 stars). 
We fit to both of the metal-poor fiducial HB sequences (M15 and M92) by minimizing the sum 
of the squares of the difference between the data and the fiducial. The best-fit distance modulus 
from HBs is 19.30 for M15 and 19.26 for M92. We calculate the associated uncertainties via jackknife 
resampling, which accounts for both the finite number of stars and the possibility of occasional
interloper stars. We compute the mean and standard deviation of the derived 
distance moduli from both methods, i.e., fitting the HB sequence (19.30, 19.26) and counting stars 
from M15 (19.250), M92 (19.125) and the isochrone (19.225). The mean is adopted as our final distance 
modulus value, and the standard deviation as our uncertainty. The distance modulus uncertainty 
of the globular clusters and the uncertainties from both jackknife resampling and our bootstrap analysis are added 
in quadrature to produce our final quoted uncertainty (see Table~\ref{tab:str}). 

We derive the distance to the other satellites by comparing their CMDs with a grid of isochrones
from both \citet{Bressan2012} and \citet{Dotter2008}, using the star counting technique described above. 
We use a 13.5 Gyr stellar population with a range of metallicities ([Fe/H]$=-2.4$, $-2.2$, $-2.0$, 
$-1.7$, $-1.5$). Similar to Sgr~II, the best-fit distance modulus values are found when the distanced-shifted 
isochrones give the maximum number of dwarf stars with $r <24$ mag. For Phe~II, which has very few RGB stars, 
we extend the limiting magnitude to $r=25$ mag to capture a larger number of main-sequence stars. The best-fit 
Dartmouth and PARSEC isochrones are the ones with [Fe/H]$=-2.4$ and [Fe/H]$=-2.0$ for Ret~II, [Fe/H]$=-2.2$ and 
[Fe/H]$=-2.0$ for Phe~II, [Fe/H]$=-2.4$ and [Fe/H]$=-2.2$ for Tuc~III, respectively. We adopt the mean of the 
results from these isochrones as our final distance moduli, and their standard deviation as our uncertainty.
The uncertainties from our bootstrap analysis are added in quadrature to produce our final quoted 
uncertainty (see Table~\ref{tab:str}). 

\subsection{Structural Properties} \label{subsec:Str}

\begin{figure*}
\includegraphics[width=0.5\linewidth]{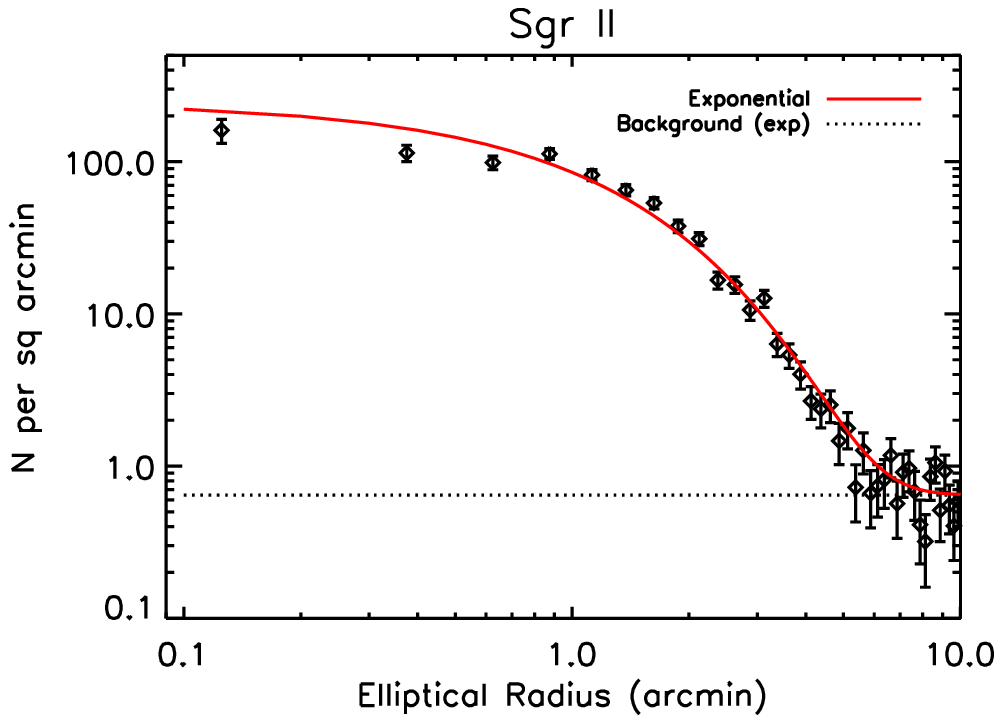}
\includegraphics[width=0.5\linewidth]{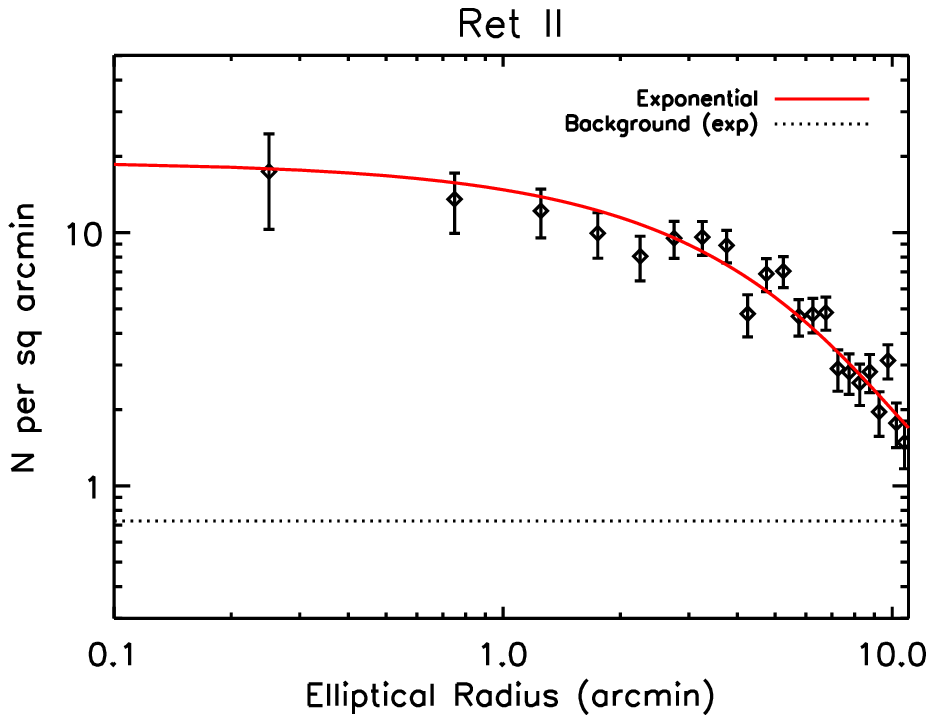}
\includegraphics[width=0.5\linewidth]{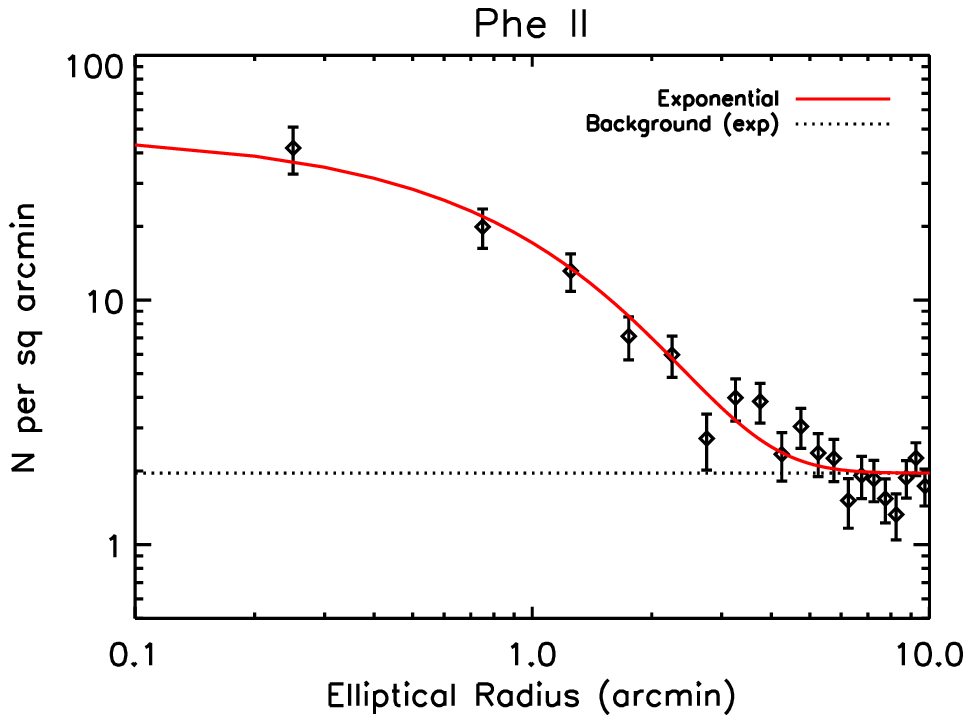}
\includegraphics[width=0.5\linewidth]{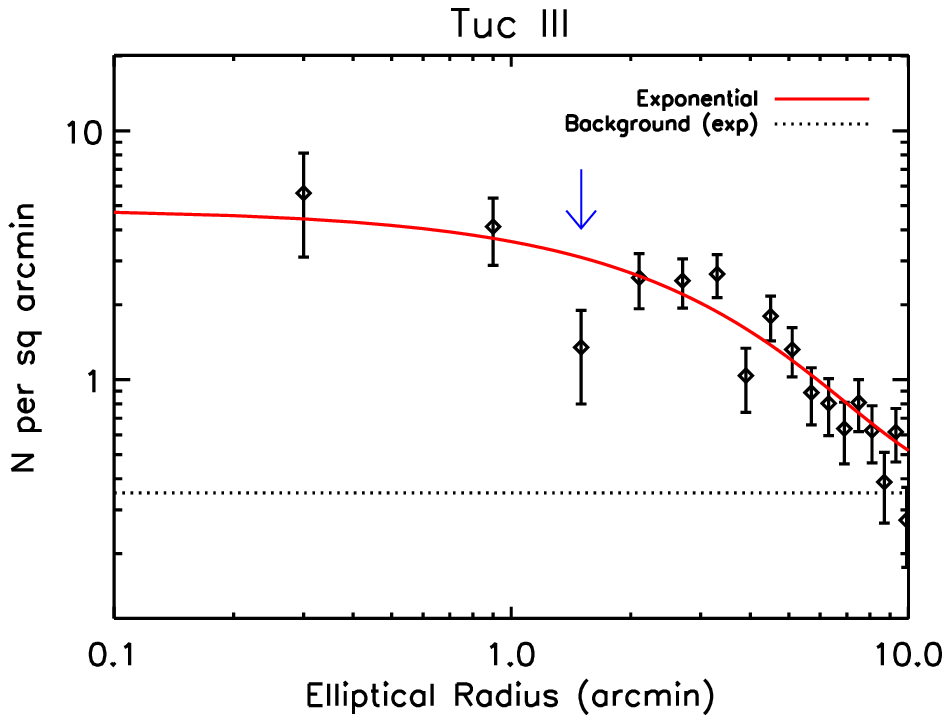}
\caption{Stellar profiles of Sgr~II, Ret~II, Phe~II, and Tuc~III. The red and dotted
lines show the best-fit one-dimensional exponential and the background surface density, found 
via our ML analysis, respectively. In Tuc~III, the blue arrow highlights the bin where
the presence of two bright stars causes a dip in the profile.
Note that our profile fits are derived from the two-dimensional 
stellar distribution, not from the one-dimensional profile.
\label{fig:surfdens}}
\end{figure*}

To constrain the structural parameters of our objects, we fit an exponential profile to the 2D distribution of stars 
consistent with each satellite by using the maximum likelihood (ML) technique of \citet{Martin2008} as implemented 
by \citet{Sand2009}. In our analysis, we only select stars consistent with the best-fitting 
Dartmouth isochrone in color-magnitude space after taking into account photometric uncertainties,
within our 90\% completeness limit (see Table~\ref{tab:obs}). We inflate the uncertainty to 0.1 mag 
when the photometric errors are $<0.1$ mag for the purpose of selecting stars to go into our ML analysis. 
For Tuc~III, a limiting magnitude of $r=24$ mag is used to avoid 
contamination from field stars and unresolved background galaxies.  
The resulting structural parameters are summarized in 
Table~\ref{tab:str}, which includes the central position, half-light radius ($r_h$), ellipticity ($\epsilon$) and 
position angle. The quoted $r_h$ is the best-fit elliptical half-light radius along the semi-major axis.
Uncertainties are determined by bootstrap resampling the data 
1000 times and recalculating the structural parameters for each resample.   
We check our results by repeating the calculations with the same set of stars, but with a limit one magnitude 
brighter. The derived structural parameters using both samples of stars are consistent within the uncertainties. 

Recently, \citet{Munoz2012} presented a suite of simulations of low-luminosity MW satellites under different 
observing conditions to recover structural parameters within 10\% or better of their true values: 
they suggested a FoV at least three times that of the half-light radius being measured, greater than 1000 
stars in the total sample, and a central density contrast of 20 over the background. These conditions are 
satisfied in our data for all but one satellite, Tuc~III. Our sample of Tuc~III has $<$500 stars and a central density 
contrast of $\sim$15; moreover Tuc III is clearly a disrupting system \citep[e.g.][and see Section~\ref{sec:tuciii}]{Drlica2015,Shipp2018} whose true structural parameters may be difficult to gauge with our ML technique.

In Figure~\ref{fig:surfdens}, we show one-dimensional stellar radial profiles, along with our best-fit 
exponential profile derived from the best-fit two-dimensional stellar distribution. 
We use elliptical bins based on the parameters from the ML analysis.
The one-dimensional representations of the exponential fit and the data are in good agreement, but we also note that parameterized models, 
condensed to one-dimension, cannot capture a satellite's potentially complex structure.
For Tuc~III, there are two bright stars which lead to incompleteness in our star counts, as shown with a blue arrow 
(see also Figure~\ref{fig:extended}). We also note that the radial density profiles of Ret~II and Tuc~III do not reach the background level while 
those of Sgr~II and Phe~II barely do. This limits our ability to investigate the outer regions of these satellites.  

\begin{table*}[th!] 
\renewcommand{\thetable}{\arabic{table}}
\centering
\small
\caption{Structural Properties} \label{tab:str}
\begin{tabular}{lcccc}
\tablewidth{0pt}
\hline
\hline
Parameter & Sgr~II & Ret~II & Phe~II & Tuc~III \\
\hline
RA (J2000.0)  		&  $19:52:39.53 \pm 3.0\arcsec$  &$03:35:47.83 \pm 24.8\arcsec$   & $23:39:58.27 \pm 8.3\arcsec$     & $23:56:25.80 \pm 36.7\arcsec$\\
DEC (J2000.0) 		&  $-22:03:54.19 \pm 2.2\arcsec$ &$-54:02:47.80 \pm 9.1\arcsec$   & $-54:24:17.83 \pm 5.7\arcsec$    & $-59:34:59.94 \pm 26.0\arcsec$\\
$m-M$ (mag)   		&  $19.2\pm0.2$                &$17.5\pm0.1$                      & $19.6\pm0.2$                     & $16.8\pm0.1$  \\
$E(B-V)$            &  $0.11$                        &$0.02$                          & $0.01$                           &  $0.01$         \\  
Distance (kpc) 		&  $70.2 \pm 5.0$                &$31.4 \pm 1.4$                  & $84.1 \pm 8.0$                   & $22.9 \pm 0.9$	\\
$N_{*}$\footnote{$N_{*}$ is the number of stars brighter than $r=24$ mag selected by the best-fit isochrone for each object.} &  1502 & 1120 & 162 & 419 \\
$M_{V}$       		&  $-5.2\pm0.1$                  & $-3.1\pm0.1$                    & $-2.7\pm0.4$                   & $-1.3\pm0.2$     \\ 
$r_{h}$ (arcmin)    &  $1.6\pm0.1$                   & $6.3\pm0.4$                    & $1.5\pm0.3$                     & $5.1\pm1.2$  \\
$r_{h}$ (pc)       	&  $32 \pm 1$                   & $58\pm4$                    & $37\pm6$                   & $34\pm8$ \\
Ellipticity         &  $<0.1$                       & $0.6\pm0.1$                     & $0.4\pm0.1$                    & $0.2\pm0.1$ \\
Position Angle (deg)  & Unconstrained                & $68\pm2$                        & $156\pm13$                     & $25\pm38$ \\
$log_{10}(J(0.5^{\circ}))$ (Ge$V^2/cm^5$) & 18.7 & 18.9 & 18.2 & 19.2 \\
\mlim\ (\msun)     &  1300    &  430   & 1400    & 210 \\
\ml\ (\mlsun)      &  0.2   & 0.4   &  2   & 0.5 \\

\hline
\end{tabular}
\end{table*}

\subsection{Absolute Magnitude}\label{subsec:Mv}

We derive absolute magnitudes for our objects by using the same procedure as in \citet{Sand2009,Sand2010}, 
as was first described in \citet{Martin2008}. First, we build a well-populated CMD (of $\sim$ 20,000 stars), 
including our completeness and photometric uncertainties, by using the best-fit Dartmouth isochrone (Sgr~II and Phe~II: [Fe/H]$=-2.2$; Ret~II and Tuc~III: [Fe/H]$=-2.4$) and its associated luminosity function with a Salpeter 
initial mass function. Then, we randomly select the same number of stars from this artificial CMD as was 
found from our exponential profile fits (over the same magnitude range as was used for the ML analysis). 
We correct the number of stars of Tuc~III for the bin obscured by bright stars, assuming that this bin 
(highlighted by blue arrow in Figure~\ref{fig:surfdens}) has a stellar density that lands on the exponential fit. 
We sum the flux of these stars, and extrapolate the flux of unaccounted stars using the adopted luminosity function.
We calculate 1000 realizations in this way, and take the mean as our absolute magnitude and its standard deviation as our 
uncertainty. To account for the distance modulus uncertainty and the uncertainty on the number of stars,
we repeat this operation 100 times, varying the presumed distance modulus and number of stars within their 
uncertainties, and use the offset from the best-fit value as the associated uncertainty. All of these error 
terms are then added in quadrature to produce our final uncertainty on the absolute magnitude.  
We note that the Dartmouth isochrone accounts for red giant branch (RGB) and main sequence stars
but not HB sequences. Adding the fluxes of our HB candidates gives the total absolute magnitude 
of $M_V=-5.2$ mag for Sgr~II, $-3.1$ mag for Ret~II and $-2.7$ mag for Phe~II, which we adopt as our final values.
For each satellite, the effect of HB candidates on the absolute magnitude is within our quoted uncertainty.
Note that Tuc~III does not have any HB candidates.

\subsection{Extended Structure Search}\label{subsec:density}

Given photometric hints that some new MW satellites may be tidally disturbed 
\citep[e.g.,][among others]{Sand2009,Munoz2010}, we search for any sign of tidal interaction, 
such as streams or other extensions, within our data. We use a matched-filter technique similar 
to \citet{Sand2012}, as was originally described in \citet{Rockosi2002}. This method maximizes 
the signal to noise in possible satellite stars over the background. 
As signal CMDs, we use the artificial ones that are created in Section~\ref{subsec:Mv} to derive the 
absolute magnitudes. For background CMDs, we use stars from a field well outside the body
of each satellite. We bin these CMDs into 0.15$\times$0.15 color-magnitude bins. 

\begin{figure*}
\includegraphics[width=0.5\linewidth]{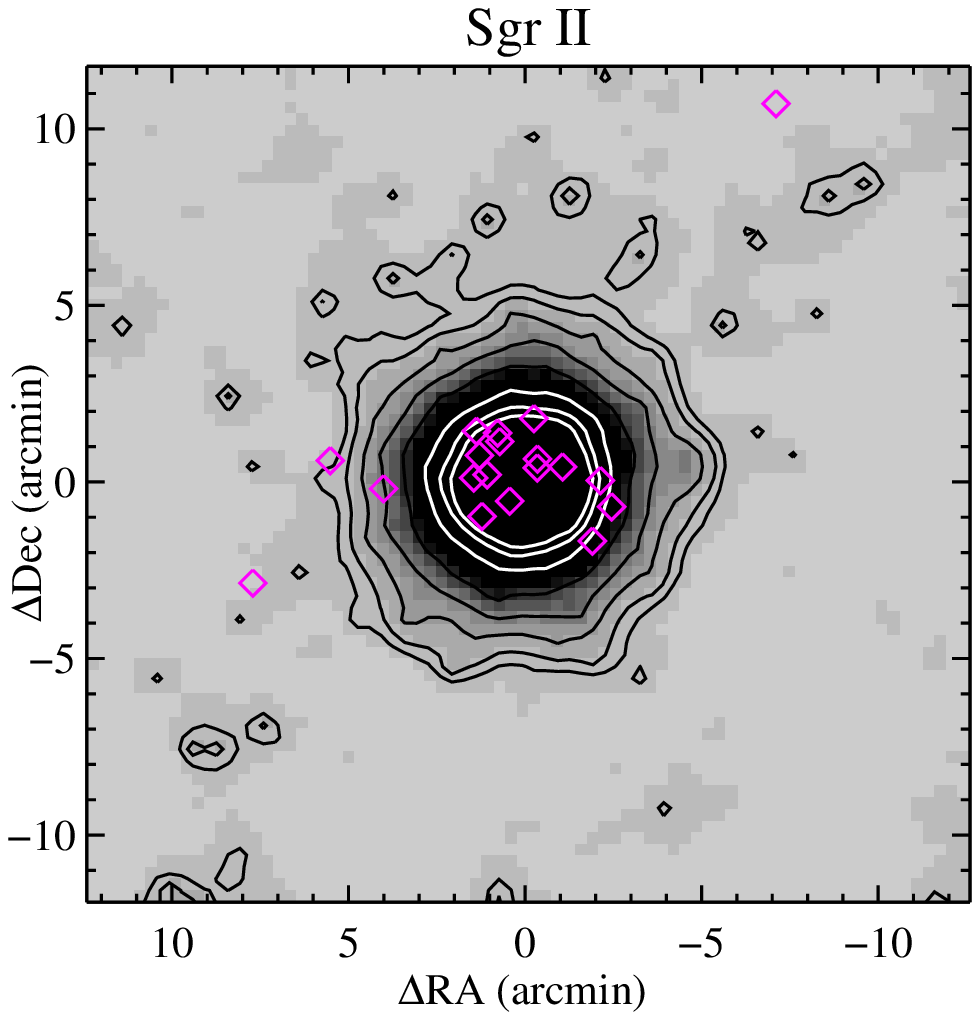}
\includegraphics[width=0.5\linewidth]{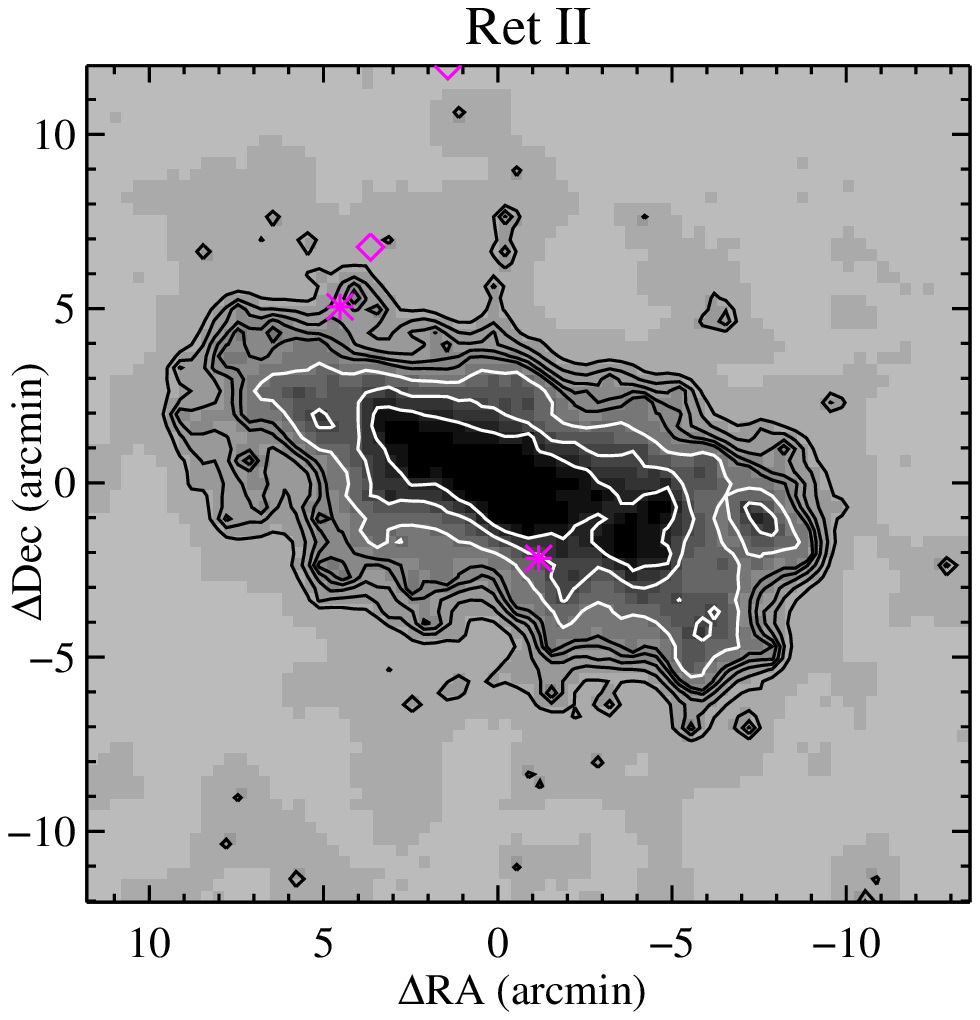}
\includegraphics[width=0.5\linewidth]{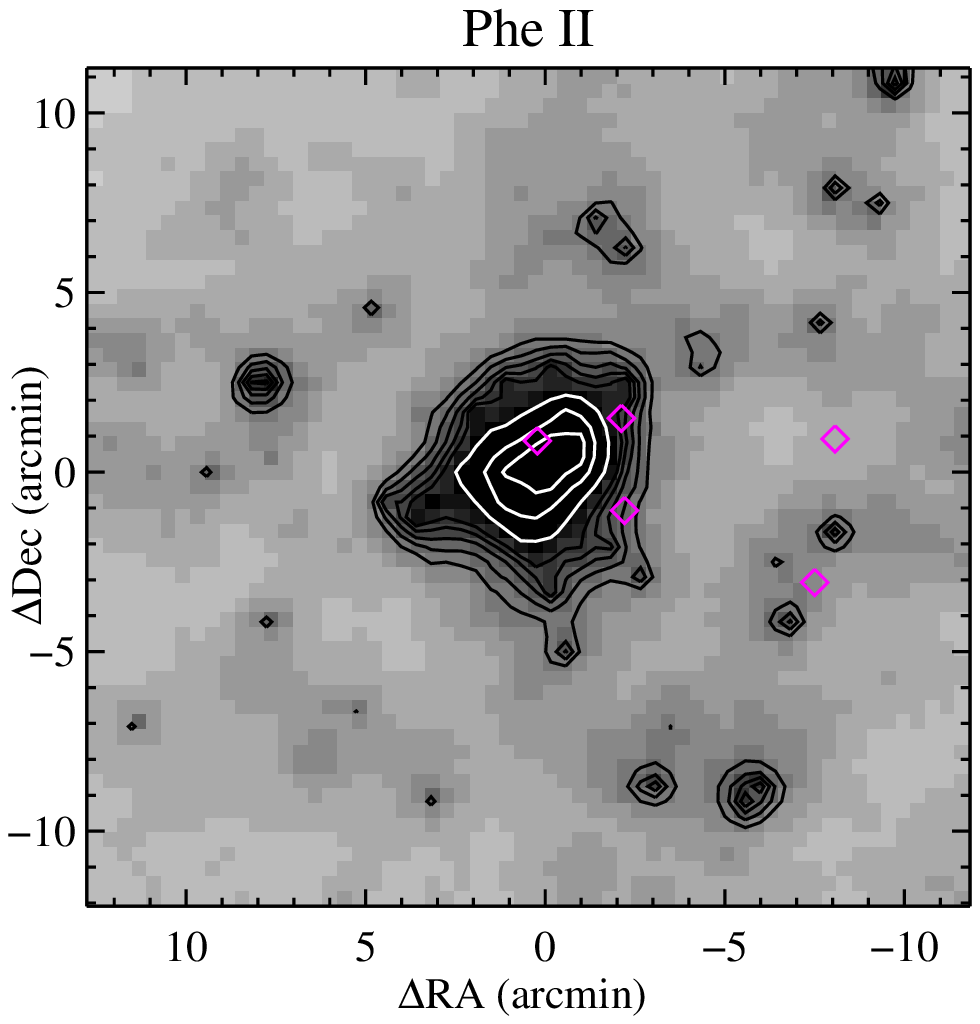}
\includegraphics[width=0.5\linewidth]{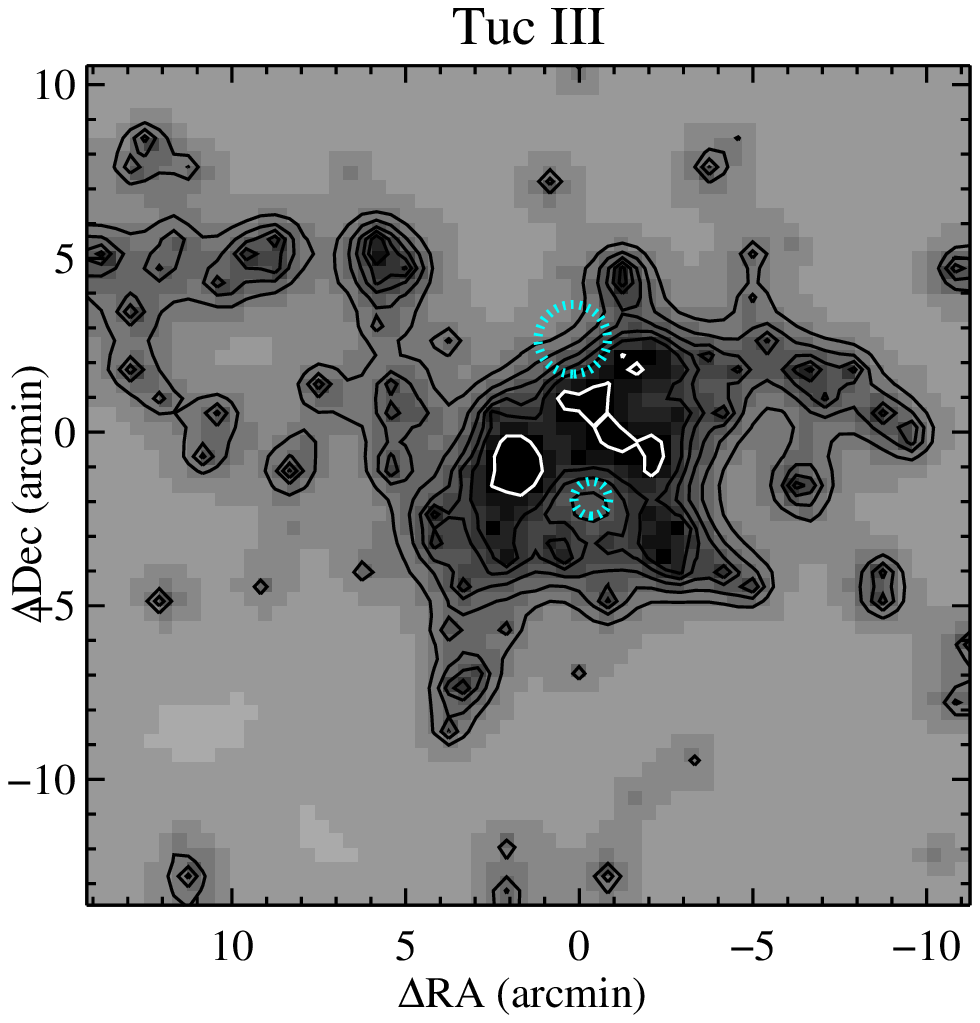}
\caption{Smoothed matched-filter maps of the satellites. The contour levels for Sgr~II show 
the 3$\sigma$, 5$\sigma$, 10$\sigma$, 20$\sigma$, 40$\sigma$, 80$\sigma$, 120$\sigma$ and 150$\sigma$ levels 
above the modal value. For the other satellites, they correspond to the 3$\sigma$, 4$\sigma$, 
5$\sigma$, 6$\sigma$, 7$\sigma$, 10$\sigma$, 15$\sigma$ and 20$\sigma$ levels. 
Magenta diamonds are likely blue HBs. In the Tuc~III map, cyan circles highlight the presence of 
bright stars which cause stellar incompleteness at those positions. 
\label{fig:extended}}
\end{figure*}

Figure~\ref{fig:extended} shows our final smoothed matched-filter maps, where we have  spatially
binned the input data, and smoothed with a Gaussian of width 1.5 times that of the pixel size. 
Spatial bins for Sgr~II and Ret~II are 20\arcsec, while for Phe~II and Tuc~III we used spatial 
pixels of 25\arcsec. The background of these smoothed maps is determined using the MMM routine in IDL. 
The main body of each satellite is clearly visible in each map. Magenta diamonds show blue HB 
candidates if there are any. In the Tuc~III map, scaled cyan circles highlight the position of bright 
stars and the approximate size of their halos -- in these regions our stellar catalogs are compromised, 
which translates to holes in our spatial maps.

\subsection{Neutral Hydrogen Content} \label{subsec:HI}

We constrain the neutral hydrogen (\HI ) content of our objects by mining the HI Parkes All-Sky Survey \citep[HIPASS,][]{Barnes2001,Kalberla2015} and Galactic All-Sky Survey \citep[GASS,][]{McClure2009} along the lines-of-sight given in Table~\ref{tab:str}. We find no evidence for \HI\ emission within the half-light radii of the objects at heliocentric velocities in the range $-400\,\kms < V_{hel} < 400~\kms$, beyond that from the \HI\ layer of the MW itself at $V_{\rm hel}\sim 0\,\kms$. We therefore place $5\sigma$ upper limits on the \HI\ mass \mlim\ of each object, using stellar radial velocity measurements to constrain the search when possible. The resulting \mlim\ and \ml\ are given in Table~\ref{tab:str}.

The measured stellar radial velocity centroids of Ret~II ($V_{sys}=[62.8\pm0.5]\,\kms$, \citealt{Simon2015}) and Tuc~III ($V_{sys}=[-102.3\pm0.4\,\mathrm{stat}\pm2.0\,\mathrm{sys}]\,\kms$, \citealt{Simon2017}) place them within and beyond the range of $V_{hel}$ contaminated by MW \HI\ emission, respectively, at the GASS sensitivity. The upper limit on the \HI\ mass \mlim\ of Ret~II derived by \citet{Westmeier2015} using HIPASS (before a $V_{sys}$ for this object was published) is only valid if it is well-separated from the MW \HI\ layer in velocity; this is clearly not the case. We therefore derive a physically meaningful \mlim\ using the GASS data (which, unlike HIPASS, accurately recover large-scale \HI\ emission) at $V_{sys}$ for Ret~II, smoothing to a spectral resolution of $10\,\kms$ and adopting this velocity width in the upper limit. Since Tuc~III is well-separated from the MW \HI\ layer, we use the more sensitive HIPASS data to derive \mlim\ in a single $26.4\,\kms$--wide Hanning-smoothed channel. 

Stellar radial velocity measurements for Sgr~II and Phe~II are not available in the literature. For Sgr~II, we compute \mlim\ from HIPASS as described for Tuc~III above; for Phe~II, we use the upper limit obtained by \citet{Westmeier2015} from reprocessed HIPASS data, adjusting \mlim\ to the distance reported in Table~6. We emphasize that these \mlim\ are valid only if $V_{sys}$ for Sgr~II and Phe~II are well-separated from MW emission along their respective lines-of-sight. Using the GASS data to identify the velocity ranges contaminated by the MW, we find that \mlim\ is valid for Sgr~II if it has $V_{sys}\lesssim -100\,\kms$ or $V_{sys}\gtrsim 125\kms$, and \mlim\ is valid for Phe~II $V_{sys}\lesssim -80\,\kms$ or $V_{sys}\gtrsim 110\kms$. 

The upper limits on the gas richness \ml\ for Sgr~II, Ret~II and Tuc~III imply that these objects are gas-poor, while \ml\ for Phe~II does not rule out the possibility of a gas reservoir similar to that of gas-rich galaxies in the Local Volume \citep{Huang2012,Bradford2015}. The general lack of \HI\ in these objects is consistent with their location within the virial radius of the MW and M31, within which all low-mass satellites are devoid of gas \citep[e.g.][]{Grcevich09,Spekkens2014,Westmeier2015}.

\section{DISCUSSION} \label{sec:disc}

\begin{figure*}[ht!]
\plotone{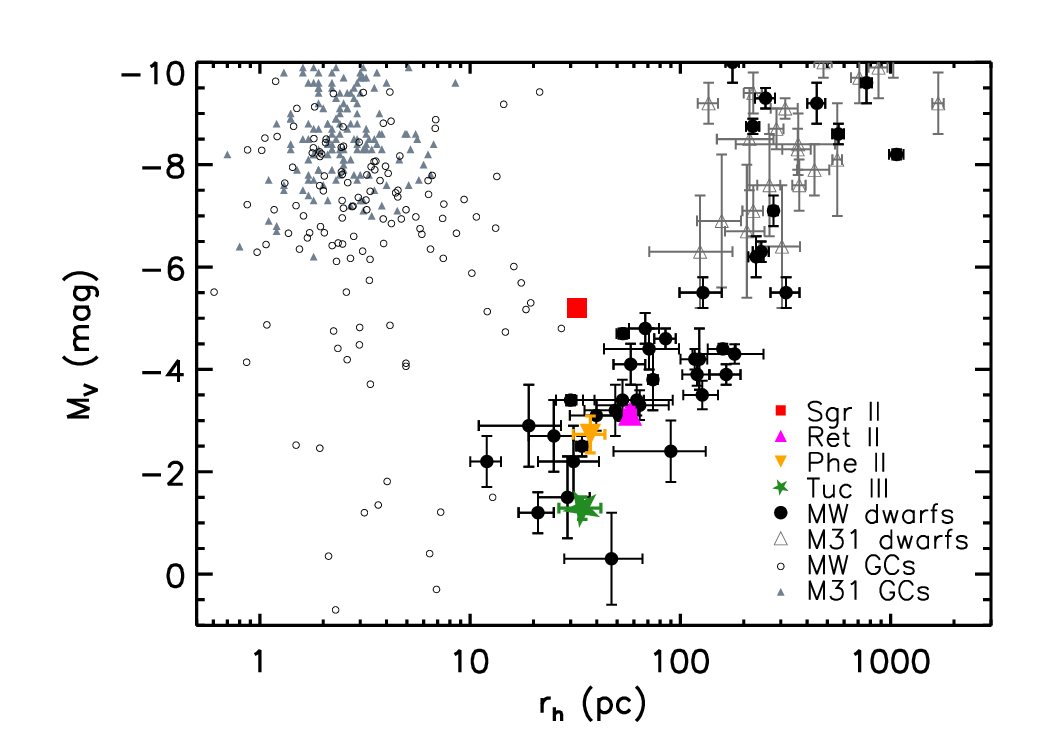}
\caption{Absolute magnitude ($M_{V}$) vs. half-light radius ($r_{h}$), showing M31 globular clusters 
(small filled gray triangles; \citealt{Strader2011}), M31 satellite galaxies (open gray triangles; \citealt{McConnachie2012}), 
MW globular clusters (small open black circles; \citealt{Harris2010,Belokurov2010,Munoz2012,Balbinot2013,KimJerjen2015,Kim2015b,Kim2016,Laevens2014}), 
and MW dwarfs (filled black circles; \citealt{McConnachie2012,Kim2016,Drlica2015,Koposov2015a,KimJerjen2015,Bechtol2015,Torrealba2016,Laevens2015,Homma2016,Homma2018,Drlica2016,Koposov18,Crnojevic2016,Carlin2017}). Our objects are displayed with a filled red square (Sgr~II), filled magenta triangle (Ret~II),
filled yellow upside down triangle (Phe~II), and filled green star (Tuc~III).
The size of the error range for Sgr~II is less than the symbol size. \label{fig:SL_plot}}
\end{figure*}

Figure~\ref{fig:SL_plot} displays the absolute magnitude ($M_{V}$) vs. half-light radius ($r_{h}$) of our objects
in the context of M31 and MW satellites. Small filled gray triangles are M31 globular clusters from \citet{Strader2011},
and open gray triangles are M31 satellite galaxies from \citet{McConnachie2012}. Small open black circles represent 
MW globular clusters from the \citet{Harris2010} catalog, supplemented by more recent discoveries \citep{Belokurov2010,Munoz2012,Balbinot2013,KimJerjen2015,Kim2015b,Kim2016,Laevens2014}. Filled black circles show
MW dwarfs from \citet{McConnachie2012}, including recently discovered MW dwarf candidates \citep{Kim2016,Drlica2015,Koposov2015a,KimJerjen2015,Bechtol2015,Torrealba2016,Laevens2015,Homma2016,Homma2018,Drlica2016,Koposov18,Carlin2017}. 
Our objects are displayed with a filled red square (Sgr~II), filled magenta triangle (Ret~II), filled yellow upside down 
triangle (Phe~II), and filled green star (Tuc~III). Sgr~II stands out with its intermediate position between the loci of dwarf 
galaxies and globular clusters in the size-luminosity plane. We discuss each satellite's derived properties in detail
in the following subsections.

\subsection{Sgr~II}

Our deep photometry provides robust constraints on the structural parameters of Sgr~II, and our results are consistent 
with the discovery analysis of \citet{Laevens2015} within the uncertainties. 
The CMD of Sgr~II (Figure~\ref{fig:cmdsag2}) has well-defined features: a narrow RGB and a clear main 
sequence turnoff (MSTO), with several blue HB candidates. In Figure~\ref{fig:extended}, 
magenta diamonds show the spatial position of the HB candidates, most of which are centrally concentrated 
around Sgr~II. Overall, the CMD is consistent with an old stellar population with [Fe/H]$\lesssim$$-$2.

Sgr~II occupies an interesting place in the size-luminosity plane (see Figure~\ref{fig:SL_plot}), 
as it sits in an intermediate position between the loci of dwarf galaxies and globular clusters. Nonetheless, 
there are several globular clusters with large half-light radii ($>10$ pc) and an absolute magnitude 
of $M_{V}>-7$ mag that occupy a similar region in parameter space  (see Figure~\ref{fig:SL_plot}). They are 
mostly distant, and the most metal poor among them (NGC~5053) has [Fe/H]$=-2.27$ \citep{Harris2010}, a value 
that is consistent with the photometric metallicity of Sgr~II.  Adding to this, Sgr~II is very round, with 
$\epsilon$$<$0.1, which is consistent with the bulk of MW GCs, although there are definite dwarf galaxies with 
similar ellipticities \citep[for instance, Leo~II; ][]{ColemanLeoII}.

The ensemble properties of Sgr~II (i.e., distance, luminosity, and size) are comparable to 
that of Palomar 14 (Pal~14), which is one of the most distant ($D=71\pm2$ kpc), faint ($M_{V}=-4.9\pm0.1$ mag), 
and diffuse ($r_{h}=46\pm3$ pc) outer Galactic halo globular clusters \citep{Sollima2011}\footnote{Note that in Figure~\ref{fig:SL_plot} we plot the properties of MW globular clusters from \citet{Harris2010}, which report $r_{h}=27$ pc and $M_{V}=-4.8$ mag for Pal~14.}. 
Similar to Sgr~II, the CMD of Pal~14 presents a narrow RGB \citep{Sollima2011}. However,
Sgr~II is more metal-poor than Pal~14 ([Fe/H]$\sim-1.5$). Based on its existing tidal tail, Pal~14 was suggested to 
be a part of a stream consisting of the Fornax dSph and globular cluster Palomar 15 \citep{Sollima2011}.

A population of extended, diffuse star clusters have recently been discovered around M31 \citep{Huxor2014} with 
roughly similar properties to Sgr~II.  Most of these systems are far from M31 itself ($>$30 kpc), and many appear 
to be associated with streams \citep{Chapman2008,Forbes2010,Mackey2010}. This supports the picture described by
\citet{Laevens2015} in which Sgr~II was brought into the MW halo along with the Sgr stream, and similar to numerous 
other MW and M31 GCs \citep[e.g.][]{Law2010,Mackey2013}. 

Despite the strong hints presented above that Sgr~II is likely an extended globular cluster, ultimately spectroscopic follow up is 
necessary to determine its true nature.

\subsection{Ret~II}\label{sec:retii}

In Figure~\ref{fig:cmdhess}, the CMD of Ret~II shows a well-defined main sequence, with several
blue HB candidates which trace its density contours. Two of our candidates 
(shown with magenta stars in Figure~\ref{fig:extended}) are the confirmed members of Ret~II \citep{Simon2015}, based on their velocities. We note that other two were not studied by \citet{Simon2015}. Our distance measurement (D=32$\pm$1 kpc) is consistent with both independent discovery analyses \citep{Bechtol2015,Koposov2015a}.
Our ML analysis suggests a similar value for its half-light radius ($r_{h}=58\pm4$ pc) to 
the result of \citet[][$r_{h}=55\pm5$ pc]{Bechtol2015}, and this is also marginally consistent with 
the result\footnote{\citet{Koposov2015a} reported the azimuthally averaged half-light radius of 32 pc. Their value 
after correcting for the ellipticity is $r_{h}=50^{+3}_{-2}$ pc} of \citet{Koposov2015a}.
Our absolute magnitude measurement ($M_{V}=-3.1\pm0.1$) is in between the results of \citet[][$M_{V}=-2.7\pm0.1$]{Koposov2015a} and \citet[][$M_{V}=-3.6\pm0.1$]{Bechtol2015}. Ret~II is among the most elongated of the MW 
satellites with $\epsilon$$\sim$0.6, comparable to Hercules ($\epsilon$$\sim$0.7) and Ursa 
Major I ($\epsilon$$\sim$0.8), both of which are at $D$$\gtrsim$100 kpc (see Table 7 of \citealt{Sand2012}).  
Therefore, Ret~II is the most elongated nearby dwarf galaxy currently known for its luminosity range.  
In spite of its elongated nature, our density map does not show any clear sign of tidal features 
within our FoV (see Figure~\ref{fig:extended}). Deep wide-field observations of Ret~II are necessary 
to truly search for signs of extended structure.  

Ret~II lands directly on the locus defined by MW ultra-faint dwarf galaxies of similar luminosity 
(see Figure~\ref{fig:SL_plot}); spectroscopic follow-up \citep{Simon2015,Walker2015,Koposov2015b} 
confirmed that it is a MW ultra-faint dwarf galaxy based on the velocities and metallicities of its stars. 
They found that Ret~II is strongly dark matter dominated and one of the most metal-poor galaxies known 
with a mean metallicity of [Fe/H]$< -2.5$. As expected from these studies, an old metal-poor Dartmouth 
isochrone ([Fe/H]$=-2.40$, [$\alpha$/Fe]$=0.40$, 13.5 Gyr) provides the best fit to the features of 
its CMD at our measured distance (see Figure~\ref{fig:cmdhess}).

A tentative gamma-ray detection associated with Ret~II \citep[at the $\sim2-3\sigma$ level;][]{Geringer2015}, 
led to excitement that a signal from annihilating dark matter was found, but this detection is not seen by all 
of the searching teams \citep[see][]{Drlica2015}. Using the photometric analysis of \citet{Bechtol2015}, 
\citet{Simon2015} conclude that Segue~1, Ursa Major~II and Coma Berenices, which possess larger J-factors, are more promising 
gamma-ray targets than Ret~II. Using our photometric analysis and the scaling relation of \citet{Pace2018}, 
we find the same value for the J-factor and confirm their conclusion.

\subsection{Phe~II}

The CMD features of Phe~II are much clearer in the Hess diagram, which accounts for the contaminating
background stars (see Figure~\ref{fig:cmdhess}). Phe~II has a sparsely populated RGB, with 
a potential population of blue HBs outside our half-light radius (see Figure~\ref{fig:extended}). 

The structural parameters of Phe~II are not well-constrained in the discovery analyses \citep{Bechtol2015,Koposov2015a}. 
Our deep photometry provides robust constraints on these parameters, which mostly agree well with the results from 
\citet{Koposov2015a}, within the uncertainties. Compared to their estimations, our ML analysis suggests a 
similar size ($r_{h}=37\pm6$ pc versus $35.7^{+8}_{-5}$\footnote{\citet{Koposov2015a} reported the azimuthally averaged half-light radius of 26 pc. This is their value after correcting for the ellipticity.} pc) and a slightly rounder shape ($\epsilon=0.4\pm0.1$ 
versus $0.5^{+0.1}_{-0.3}$) with similar luminosity ($M_{V}=-2.7\pm0.4$ versus $-2.8\pm0.2$) for Phe~II. 
We note our absolute magnitude is fainter than that in \citet{Bechtol2015} ($M_{V} = -3.7\pm0.4$), however
their value would be $M_{V} = -3.4\pm0.4$ if shifted by our distance modulus ($m-M = 19.6$).
In Figure~\ref{fig:SL_plot}, it lands directly on the locus defined by other MW ultra-faint 
dwarf galaxies of similar luminosity. 
Judging from the ellipticity and position in the size-luminosity plane, it is likely that 
Phe~II is a dwarf galaxy.

\subsection{Tuc~III}\label{sec:tuciii}

Our deep photometry of Tuc~III reveals a well-defined narrow main sequence consistent with old, metal-poor 
stellar populations, as expected from the spectroscopic metallicity measurement
([Fe/H]$=-2.42^{+0.07}_{-0.08}$, \citealt{Simon2017}).
There are no HB candidates within our FoV, both within our Magellan photometric catalog 
(which saturates at $r$$\approx$18 mag) and the DES catalog, which goes to brighter magnitudes \citep{Drlica2015}. 

Tuc~III is known to host a stellar stream extending at least $\pm2^{\circ}$ from its 
core \citep{Drlica2015,Shipp2018}. Despite our narrow FoV, and the likelihood that 
it is contaminated with Tuc~III stars throughout, there is still evidence for the stream 
in our stellar density map (see Figure~\ref{fig:extended}).
The presence of bright stars in our field (cyan circles) causes some irregularities 
in the contours of the satellite body near its center. Despite the existence of a stellar 
stream, \citet{Drlica2015} suggest a relatively low ellipticity for Tuc~III. Our ML analysis 
gives an ellipticity of $\epsilon=0.2\pm0.1$ with a major-axis position angle of $\sim25^{\circ}$. 
Compared to the discovery analysis ($r_{h}=44\pm6$ pc, $M_{V}=-2.4\pm0.4$),
our photometry suggests a similar size ($r_{h}=34\pm8$ pc) within the uncertainties, but fainter absolute magnitude ($M_{V}=-1.3\pm0.2$). 
The stellar stream together with the narrow FoV and presence of bright stars makes
it hard to obtain a robust estimate of Tuc III's structural parameters, therefore our results should be used with caution.  

Spectroscopic follow-up \citep{Simon2017} has been unable to conclusively determine Tuc~III's dynamical status 
and dark matter content. Compared to any known dwarf galaxy, they found a smaller metallicity dispersion 
and likely a smaller velocity dispersion for Tuc~III, and tentatively suggested that it is the tidally-stripped 
remnant of a dark matter-dominated dwarf galaxy. 
Moreover, its location in the size-luminosity plane favors a ultra-faint dwarf galaxy interpretation (see Figure~\ref{fig:SL_plot}), comparable to Segue~I ($M_{V}=-1.5\pm0.8$, $r_{h}=29\pm8$ pc) and Triangulum~II 
($M_{V}=-1.2\pm0.4$, $r_{h}=21\pm4$ pc), both of which have been suggested to be tidally stripped 
\citep[e.g.,][]{Niederste2009,Kirby2017}. 

\section{SUMMARY AND CONCLUSIONS}\label{sec:conc}
We have presented deep Magellan/Megacam photometric follow-up observations of new MW satellites Sgr~II, 
Ret~II, Phe~II, and Tuc~III. Our photometry reaches $\sim$2-3 mag deeper than the original discovery data 
for each object, and allows us to revisit the distance, structural properties, and luminosity measurements.  
An archival analysis allowed us to place HI gas mass limits on each system, which are all devoid of gas 
as expected for their location within the virial radius of the MW \citep[e.g.][]{Spekkens2014}.

Sgr~II stands in an interesting location in the size-luminosity plane, just between the loci of 
dwarf galaxies and globular clusters. However, the ensemble of its structural parameters is more 
consistent with a globular cluster classification. In particular, many of its physical properties are
comparable to those of Pal~14. Spectroscopic follow up is necessary to determine its true nature.

Two independent discovery analyses found different values for the structural parameters of Ret~II and Phe~II. 
Our deep photometry resolves this inconsistency, and provides robust constraints on these parameters.
We find $M_{V}=-3.1\pm0.1$ for Ret~II and $M_{V}=-2.7\pm0.4$ for Phe~II, with corresponding 
half-light radii of $r_{h,RetII}=58\pm4$ pc and $r_{h,PheII}=37\pm6$ pc. Ret~II and Phe~II therefore land directly 
on the locus defined by MW ultra-faint dwarf galaxies of similar luminosity. Ret~II is the 
most elongated nearby dwarf galaxy currently known for its luminosity range, and it is more likely that Phe~II 
is a dwarf galaxy than a star cluster. 

Tuc~III is extremely faint with $M_{V}=-1.3\pm0.2$, and compact ($r_{h,TucIII}=34\pm8$ pc).
It is apparently made up of an old, metal-poor stellar population, as expected from its measured spectroscopic 
metallicity of [Fe/H]$=-2.42^{+0.07}_{-0.08}$ \citep{Simon2017}. Our photometry suggests that Tuc~III is a 
tidally-disrupted dwarf galaxy. 

Finally, we search for any clear sign of extended structure for these satellites (see Figure~\ref{fig:extended}). 
We find no evidence for structural anomalies or tidal disruption in Sgr~II and Phe~II. In spite of its apparent 
high ellipticity, Ret~II also does not show any firm evidence of extra-tidal material outside
the satellite. The stellar density map of Tuc~III is of particular interest, because it is known to be associated 
with a stellar stream. Our deep imaging allows us to map the connection between the stellar steam and the body of Tuc~III. 
However, deep wide-field observations, in particular for Ret~II and Tuc~III, are necessary to definitively investigate the 
outer regions of these systems.

\acknowledgments
Research by DJS and BM is supported by NSF grants AST-1412504 and AST-1517649.
We thank Keith Bechtol for helpful discussions. We  also  wish  to  thank  the  anonymous
referee whose comments improved the content of this paper. This paper includes data gathered 
with the 6.5~m Magellan Telescopes located at Las Campanas Observatory, Chile. 
This paper uses data products produced by the OIR Telescope Data Center, 
supported by the Smithsonian Astrophysical Observatory.

This project used public archival data from the Pan-STARRS1 Surveys, which have been 
made possible through contributions of the Institute for Astronomy, the University of 
Hawaii, the Pan-STARRS Project Office, the Max-Planck Society and its participating 
institutes, the Max Planck Institute for Astronomy, Heidelberg and the Max Planck Institute 
for Extraterrestrial 
Physics, Garching, The Johns Hopkins University, Durham University, the University of
Edinburgh, Queen's University Belfast, the Harvard-Smithsonian Center for Astrophysics, 
the Las Cumbres Observatory Global Telescope Network Incorporated, the National Central
University of Taiwan, the Space Telescope Science Institute, the National Aeronautics 
and Space Administration under Grant No. NNX08AR22G issued through the Planetary Science 
Division of the NASA Science Mission Directorate, the National Science Foundation under 
Grant AST-1238877, the University of Maryland, Eotvos Lorand University (ELTE), and 
the Los Alamos National Laboratory.

This project used public archival data from the Dark Energy Survey (DES). Funding for 
the DES Projects has been provided by the U.S. Department of Energy, the U.S. National 
Science Foundation, the Ministry of Science and Education of Spain, the Science and 
Technology Facilities Council of the United Kingdom, the Higher Education Funding Council
for England, the National Center for Supercomputing Applications at the University of Illinois
at Urbana-Champaign, the Kavli Institute of Cosmological Physics at the University of Chicago, 
the Center for Cosmology and Astro-Particle Physics at the Ohio State University, the Mitchell
Institute for Fundamental Physics and Astronomy at Texas A\&M University, Financiadora de 
Estudos e Projetos, Funda\c{c}\~{a}o Carlos Chagas Filho de Amparo \`{a} Pesquisa do Estado do Rio 
de Janeiro, Conselho Nacional de Desenvolvimento Cient\'{i}fico e Tecnol\'{o}gico and the Minist\'{e}rio
da Ci\^{e}ncia, Tecnologia e Inova\c{c}\~{a}o, the Deutsche Forschungsgemeinschaft and the Collaborating 
Institutions in the Dark Energy Survey. The Collaborating Institutions are Argonne National 
Laboratory, the University of California
at Santa Cruz, the University of Cambridge, Centro de Investigaciones En\'{e}rgeticas, 
Medioambientales y Tecnol\'{o}gicas-Madrid, the University of Chicago, University College London, 
the DES-Brazil Consortium, the University of Edinburgh, the Eidgen\"{o}ssische Technische Hochschule 
(ETH) Z\"{u}rich, Fermi National Accelerator Laboratory, the University of Illinois at Urbana-Champaign, 
the Institut de Ci\`{e}ncies de l’Espai (IEEC/CSIC), the Institut de F\'{i}sica d'Altes Energies, Lawrence 
Berkeley National Laboratory, the Ludwig-Maximilians Universit\"{a}t M\"{u}nchen and the associated Excellence 
Cluster Universe, the University of Michigan, the National Optical Astronomy Observatory, the University 
of Nottingham, The Ohio State University, the OzDES Membership Consortium, the University of Pennsylvania, 
the University of Portsmouth, SLAC National Accelerator Laboratory, Stanford University, the University 
of Sussex, and Texas A\&M University. Based in part on observations at Cerro Tololo Inter-American 
Observatory, National Optical Astronomy Observatory, which is operated by the Association of 
Universities for Research in Astronomy (AURA) under 
a cooperative agreement with the National Science Foundation.

\vspace{5mm}
\facilities{Las Campanas Observatory: Magellan Clay Telescope/Megacam}


\software{SExtractor \citep{Bertin1996}, IDL astronomy users library \citep{Landsman1993}, NumPy \citep{VanDerWalt2011}}

\bibliographystyle{aasjournal}
\bibliography{reference}





\end{document}